\documentclass[12pt]{article} 

\usepackage[english]{babel}
\usepackage {graphicx}
\usepackage {graphics}
\usepackage{epsfig}
\usepackage {floatflt}

\usepackage {caption3}
\usepackage[labelsep=period]{caption}[2007/11/04 v.3.1e]

\usepackage {amssymb}

\begin{document}

\noindent
{\bf UDC} 530.12;$\,$ 531.51
\vspace{0.5cm}

\begin{center}
{\large\bf AN INTERIOR SPHERICAL STATIC SOLUTION OF EINSTEIN-MAXWELL EQUATIONS}\footnote
{ Translation from Russian:
 Baranov A.M. An interior spherical static solution of Einstein-Maxwell equations, Krasnoyarsk State University, Krasnoyarsk, 1973, 9 p. Deposited in VINITI USSR June 5, 1973, no.6729-73~Dep. (in Russian). Original article submitted in June 5, 1973.}
\end{center}

\smallskip
\begin{center}
{\large\bf A.M.Baranov}{\footnote[2]{ Krasnoyarsk State Pedagogical University named after V.P.Astaf'ev (KSPU), \newline 
Krasnoyarsk, Russia.}$^{,}$ 
\footnote[3]{Siberian State University of Science and Technologies named after acad. M. F. Reshetnev  (SibSU), Krasnoyarsk, Russia.}$^{,}$
\footnote[4] {E-mail: alex\_m\_bar@mail.ru; ambar\_@kspu.ru} }
\end{center}

\smallskip
\begin{quote}
{\small\noindent
The exact interior static solution of Einstein-Maxwell equations in Bondi's coordinates for the electrically charged fluid ball is obtained as a source of the Reissner-Nordstr\"om solution.}
\end{quote}

\vspace{0.3cm}


\vspace{0.5cm}

\vspace{1.0cm}

\noindent
The Reissner-Nordstr\"om solution is well-known as an exterior solution of the Einstein-Maxwell equations with an electric charge \cite{Mitsk1}. Therefore it would be naturally to find an interior solution of a charged substance as an analog of Schwarzschild's interior solution \cite{Synge}.

Brahmachary \cite{Br1, Br2} 
obtained an interior solution as a combination of both solutions 
for a charged core without the substance and  an uncharged spherical distribution of the matter around the core.
Such description is not physically satisfactory. Therefore in present paper an attempt of finding of a static solution for the Einstein-Maxwell equations with a spherical distribution of a charged perfect fluid was done.

The method of $\tau$-field for an introduction of physically observable values is used here after \cite{Zel, Mitsk2}. A physically observable strength of an electrical field and a charge density respectively can be written according to this formalism in a comoving frame of reference as

$$
E_{\nu}=\displaystyle\frac{H_{0 \nu}}{\sqrt{g_{00}}};
\eqno{(1)}
$$

$$
\rho = j^{0} \sqrt{g_{00}},
\eqno{(2)}
$$

\noindent
where $H_{\mu \nu}$ is a tensor of an electromagnetic field; $\;g_{00}$ is a $00$-component of a metric tensor; $\;j^{0}$ is a $0$-component of an electrical 4-current density; (the greek subscripts run through the values $0,\,1,\,2,\,3$).

Further we shall write a square of the time-space interval using Bondi's coordinates as

$$
ds^2 = F du^2 + 2D dudr - r^2 (d\theta^2 + sin^2\theta \,d\varphi^2),
\eqno{(3)}
$$
\noindent 
where $F=g_{00};\,$ $ D=g_{01},\;$ $u$ is a retarded time coordinate; $r$ is a radial coordinate and $\varphi, \theta$ are angle variables. The light velocity is here equal to an unit. 

The Maxwell equations \cite{Dau},

$$
H^{\mu \nu}_{\;\;\;\; ;\,\nu} = \displaystyle\frac{1}{\sqrt{-g}} \left(\sqrt{-g}H^{\mu \nu}\right)_{,\,\nu} = -4\pi j^{\mu}, 
\eqno{(4)}
$$
\noindent
for spherical and static conditions can be rewritten in a form

$$
\left(\displaystyle\frac{r^2 H_{01}}{D}\right)_{,\,r} = 4\pi r^2 \rho D F^{-1/2},
\eqno{(5)}
$$
\noindent
where $g$ is a metric tensor determinant; semicolon marks a covariant deriva-\linebreak
tive; comma marks a private derivative.

The substance energy-momentum tensor without a viscosity is chosen in form 
$$
\left(T_{\mu \nu}\right)_{matter} = (p+\mu)u^{\mu}u^{\nu}-p\, g_{\mu \nu},
\eqno{(6)}
$$

\noindent
where $p$ is a pressure; $\mu$ is a mass density; $u^{\mu} = \displaystyle\frac{dx^{\mu}}{ds}.$

To obtain a solution in terms of elementary functions we choose a function of mass density as 

$$
\mu(r) = \beta - \alpha\,r^2,
\eqno{(7)}
$$

\noindent
where $\beta$ and $\alpha$ are constants.

For a physically observable density of an electrical field's energy in a charged substance a following functional dependence is supposed

$$
W_{el}=-\displaystyle\frac{1}{8\pi} g^{\nu \mu} E_{\nu}E_{\mu} = -\displaystyle\frac{1}{8\pi} H_{0 \nu}H_{0 \mu} = \displaystyle\frac{1}{8\pi} \lambda r^2,
\eqno{(8)}
$$

\noindent 
where $ \lambda =const > 0.$

Now we shall rewrite the Einstein-Maxwell system of equations as

$$
\displaystyle\frac{F}{rD}\left( ln D\right)_{,\,r} = \displaystyle\frac{\varkappa}{2} (p+\mu); 
\eqno{(9.a)}
$$

$$
\displaystyle{\frac{F}{rD^2}\left(ln D\right)_{,\,r}-\frac{1}{2D^2}(F_{,\,r,\,r}+\frac{2}{r}F_{,\,r} - F_{,\,r}\left(ln D\right)_{,\,r} ) = -\varkappa\left(p + \frac{1}{8\pi}\frac{E^2}{D^2}\right) }; 
\eqno{(9.b)}
$$

$$
\displaystyle{\frac{1}{r^2 D^2}(-D^2+F+r F_{,\,r} - r F\left(ln D\right)_{,\,r} ) = -\varkappa\left(\frac{1}{2}(\mu - p) + \frac{1}{8\pi}\frac{E^2}{D^2}\right) }; 
\eqno{(9.c)}
$$

$$
\displaystyle{\left(\frac{r^2 F}{D}\right)_{,\,r} = 4\pi r^2 \rho D F^{-\frac{1}{2}}}; 
\eqno{(9.d)}
$$

\noindent 
where $\varkappa = 8\pi;$ Newton's gravitational constant $\gamma$ is here equal to unit; $E = H_{0\,1}$ is a coordinate strength of the electric field.

Now we take 

$$
D(r)=\left(F(r)\,{\varepsilon(r)}^{-1}\right)^{\frac{1}{2}}, 
\eqno{(10)}
$$
\noindent 
where $\varepsilon(r) = 1+\varphi(r),\;$ $\varphi(r)$ is an unknown function for now.

The formula (10) is suggested by known transformation from coordinates of curvatures to Bondi's coordinates for Schwarzschild's interior solution.

The junction conditions with an exterior solution of Reissner-Nordsrem can be written as 

$$
F(r)=1-\displaystyle{\frac{2m}{a}+\frac{q^2}{a^2}}; \,\,\,\, {F(r)}_{,\,r}=\displaystyle{\frac{2m}{a^2}(m-\frac{q^2}{a}) };
\eqno{(11.a)}
$$

$$
W_{el}=\displaystyle{\frac{q^2}{8\pi a^4}};
\eqno{(11.b)}
$$

$$
p(a)=0; \;\;\; D(a) = 1; \;\;\; D_{,\,r}(a) = 0, 
\eqno{(11.c)}
$$
\noindent
where $q,\; m$ are an integral electric charge and a mass of the ball respectively; $a$ is a radius of ball.

In general the condition $D_{,\,r}(a) = 0$ can be not correct in chosen coordina-\linebreak
tes \cite{Synge, Den}.

Now we go over to equations (9). If we use equations (10), (8) and (7), we obtain from (9.c) new equation. A quadrature of this equation is equal to

$$
\varphi(r)=-\eta r^2 + \xi r^4,
\eqno{(12)}
$$

\vspace{2mm}

\noindent
where $\eta = \displaystyle\frac{1}{3} \varkappa \beta\;$ and $\xi = \displaystyle\frac{\varkappa}{5} (\alpha - \displaystyle\frac{\lambda^2}{8\pi})$ are parameters.

\vspace{2mm}
Further we subtract the equation (9.b) from (9.a) and next subtract (9.c). Taking into account the equation (9.a), we get equation

$$
\displaystyle{\frac{F_{,\,r,\,r}}{2F} \varepsilon + \frac{F_{,\,r}}{F}\left(\frac{\varepsilon_{,\,r}}{2} - \frac{\varepsilon}{r} \right) -\frac{1}{4} \left(\frac{F,\,r}{F} \right)^2 \varepsilon + \frac{1}{r^2} - \frac{\varepsilon}{r^2} + \frac{\varepsilon_{,\,r}}{2r} = \frac{\varkappa}{8\pi} \lambda^2 r^2}.
\eqno{(13)}
$$

Next we make substitution

$$
U(r)= \displaystyle{(ln F)_{,\,r} \,\varepsilon^{\frac{1}{2}}\, r^{-1} }.
\eqno{(14)}
$$

Finally, taking into account Eq.(12), we obtain Riccati's general
equa-\linebreak 
tion \cite{Kam}

$$
U_{,\,r} + \displaystyle{\frac{1}{2}r \varepsilon^{-\frac{1}{2}}\, U^2 = 2\nu\,r\, \varepsilon^{-\frac{1}{2}}},
\eqno{(15)}
$$
\noindent
where a parameter $\nu ={\displaystyle\frac{\varkappa}{4\pi}} \lambda^{2} - \xi.$

The solution to the equation (15)  strongly depends  on variability of signs of $\nu$ and $\xi$, i.e. from a relation between parameters $\alpha$ and $\lambda.$ 
Here we will consider an opportunity only with 

$$
0 < \alpha <\displaystyle\frac{\lambda^2}{8\pi},\;\; \hbox{then} \;\; -\xi <0; \;\; \nu >0.
\eqno{(16)}
$$

In this case the separation of variables may be made in the equation (15) and after the integration we have function 

$$
U(r) = 2\sqrt{\nu} \displaystyle{\frac{R(a)+\displaystyle\frac{P}{Q} R(r)}{R(a)-\displaystyle\frac{P}{Q} R(r)} },
\eqno{(17)}
$$
\noindent 
where

$$
P = \eta - \varkappa\,\alpha\,a^2 - 4\xi\,a^2 - 2\sqrt{\nu\,\varepsilon(a)} ;
$$
\vspace{0.2cm}
$$
Q = \eta - \varkappa\,\alpha\,a^2 - 4\xi\,a^2 + 2\sqrt{\nu\,\varepsilon(a)} ;
$$
\vspace{0.2cm}
$$
R(r) = exp{\left(-\sqrt{\frac{\nu}{\xi}}\, \arcsin{\left(\frac{\xi\,r^2 + \eta/2}{ \sqrt{\xi+\eta^2/4}}  \right)}\right)}.
$$
\vspace{0.1cm}

The constant of integration is found here  for the true junction condition of pressure (11.c). After that from (14) and (17) we find

$$
F(r) = \varepsilon(a)\,R(a){R(r)}^{-1} \displaystyle{\left(1 - \frac{P}{Q}\frac{R(r)}{R(a)}\right)^{2}\left(1-\frac{P}{Q}\right)^{-2} }.
\eqno{(18)}
$$
\noindent
The condition $D(a)=1$ was here used to find the integration constant.

$$
D(r) = \left(\varepsilon(a)\,R(a)\right)^{1/2}\left(\varepsilon(r)R(r)\right)^{-1/2} \displaystyle{\left(1 - \frac{P}{Q}\frac{R(r)}{R(a)}\right)\left(1-\frac{P}{Q}\right)^{-1} }.
\eqno{(19)}
$$

We find the pressure from (9.a) as
$$
p(r) = -\frac{1}{3}\beta + \alpha r^2 +4\frac{\xi}{\varkappa} r^2 + \frac{2}{\varkappa} (\nu \varepsilon(r))^{1/2}\frac{\left(1 - \displaystyle\frac{P}{Q}\frac{R(r)}{R(a)}\right)}
{\left(1-\displaystyle\frac{P}{Q}\right)}.
\eqno{(20)}
$$

This solution (see Eqs. (19), (20), (18)) )) goes over into the standard solution of Schwarzschild for the perfect fluid if at first we shall take limit $\alpha \rightarrow 0,$ and then go to the limit procedure  $q \rightarrow 0$ with $\beta = \mu_0 = const.$ As result we shall have ( see e.g. \cite{Synge})

$$
\lim_{q \rightarrow 0} \lim_{\alpha \rightarrow 0}{F(r)} = \left(\frac{3 \sqrt{1-\eta a^2} - \sqrt{1-\eta r^2} }{2}\right)^2 ;
$$
$$
\lim_{q \rightarrow 0} \lim_{\alpha \rightarrow 0}{D(r)} = \frac{1}{2} \left(\frac{3 \sqrt{1-\eta a^2}}{\sqrt{1-\eta r^2}} - 1\right);
\eqno{(21)}
$$
$$
\lim_{q \rightarrow 0} \lim_{\alpha \rightarrow 0}{p(r)} = \mu_0 \left(\frac{\sqrt{1-\eta r^2} - \sqrt{1-\eta a^2} }{3 \sqrt{1-\eta a^2} - \sqrt{1-\eta r^2}}\right),
$$
\noindent
where $\eta = \displaystyle\frac{1}{3}\varkappa \mu_0.$

We will take $\mu(a)=0$ to find  $\alpha$ and  $\beta$  out of Eq.(11.a) in the form 

$$
\alpha = \frac{15}{\varkappa a^3}\left(m -\frac{3 q^2}{5 a}\right).
\eqno{(22)}
$$
\noindent
The condition $D_{,r}(a) =0$ is here the true.

Further we find a restriction on an integral mass (in dimensional units), using (16) and $\lambda^2 = \displaystyle\frac{q^2}{a^6}$ out of (11.b), 

$$
\frac{3}{5} \frac{q^2}{a c^2} < m < \frac{2}{3} \frac{q^2}{a c^2}.
\eqno{(23)}
$$

\vspace{2mm}
We have physically observable density of charge after using (9.d), (10), (8), 

$$
\rho(r) = \displaystyle{\frac{q}{\left(\displaystyle\frac{4}{3}\pi a^3\right)}}\cdot \varepsilon^{1/2}(r).
\eqno{(24)}
$$

The function $\rho(r)$ is monotonic in range $0 \leq r \leq a$  and gets maximum 
at $r=0.$

We must note, that at $r=0$ the mass density must be positive. Then simple demand follows from (23) and (7): 
$$
m > \frac{1}{2} \frac{q^2}{a c^2} .
\eqno{(25)}
$$
This demand is satisfied for the expression (24). 

Herewith the observable strength of n electrical field equals to (in view of (1), (8))
$$
E_1(r) = \frac{4}{3} \pi \rho r.
\eqno{(26)}
$$

We see that $E_1(1)$ is simple generalization of an expression for  electrical field strength of  homogeneously charged  ball in a "flat" case, i.e. for case with the zero gravitational constant.

The monotonicity demand of the function $E_1(r)$ inside of ball leads to   lower limit of the ball radius 
$$
a > \frac{m \gamma}{c^2} = R_g,
\eqno{(27)}
$$
\noindent
where $R_g$ is radius of Schawzcshild.

\vspace{10mm}

\section*{Acknowledgements} 

The author expresses his sincere gratitude to Professor N.V.Mitskievich for the useful and fruitful discussion of the problem.

\end{document}